# Title: Optical Control of Integer and Fractional Chern Insulators

**Authors:** William Holtzmann[1*], Weijie Li[1*], Eric Anderson[1], Jiaqi Cai[1], Heonjoon Park[1], Chaowei Hu[1], Takashi Taniguchi[2], Kenji Watanabe[3], Jiun-Haw Chu[1], Di Xiao[4,1], Ting Cao[4], and Xiaodong Xu[1,4#]

[1]Department of Physics, University of Washington, Seattle, Washington 98195, USA
[2]Research Center for Materials Nanoarchitectonics, National Institute for Materials Science, 1-1 Namiki, Tsukuba 305-0044, Japan
[3]Research Center for Electronic and Optical Materials, National Institute for Materials Science, 1-1 Namiki, Tsukuba 305-0044, Japan
[4]Department of Materials Science and Engineering, University of Washington, Seattle, Washington 98195, USA

[*] These authors contribute equally to this work.

[#]Corresponding author's email: xuxd@uw.edu

**Abstract:** Optical control of topology, particularly in the presence of electron correlations, is a fascinating topic with broad scientific and technological impact. Twisted MoTe$_2$ bilayer (tMoTe$_2$) is a newly discovered zero-field fractional Chern insulator (FCI), exhibiting the fractionally quantized anomalous Hall (FQAH) effect. Since the chirality of the edge states and sign of the Chern number are determined by the underlying ferromagnetic polarization, manipulation of ferromagnetism would realize control of the CI/FCI states. Here, we demonstrate control and switching of ferromagnetic polarization, and thus the CI and FCI states by circularly polarized optical pumping in tMoTe$_2$. At low optical excitation power, we achieve on-demand preparation of ferromagnetic polarization by optical training, *i.e.*, electrically tuning the system from non-ferromagnetic to desirable ferromagnetic states accompanied with helicity-selective optical pumping. With increased excitation power, we further realize direct optical switching of ferromagnetic polarization at a temperature far below the Curie temperature. Both optical training and direct switching of ferromagnetism are most effective near CI/FCI states, which we attribute to a gap enhanced valley polarization of photo-injected holes. We show that the magnetization can be dynamically switched by modulating the helicity of optical excitation. Spatially resolved measurements further demonstrate optical writing of a ferromagnetic, and thus a CI (or FCI) domain. Our work realizes precise optical control of a topological quantum many-body system with potential applications in topological spintronics, quantum memories, and creation of exotic edge states by programmable patterning of integer and fractional QAH domains.

**Main Text**

Optical excitation is a powerful means to access, create, and manipulate quantum phases of matter. An emergent topic is the usage of light to control quantum geometry and topology of solids[1–4]. Recent examples include optical training of magnetism in the topological magnet $MnBi_2Te_4$[5], light induced anomalous Hall effect in graphene[6,7], and optically driven topological phase transitions in a range of topological insulators and semimetals[8-10]. An outstanding challenge in the field, however, is optical control of strongly correlated topological phase matter, such as systems hosting the fractional quantum Hall effect. This would enable new approaches to create and manipulate anyonic quasiparticle excitations[11–13] for potential applications in transduction and fault-tolerant topological quantum computation[14]. The recent discovery of the fractional quantum anomalous Hall (FQAH) effect in twisted $MoTe_2$ bilayer ($tMoTe_2$)[15-22] – a zero-field fractional Chern insulator (FCI)[23–28], offers an outstanding platform to explore the interplay between light and correlated topology.

There are several unique properties of $tMoTe_2$ which are critical for achieving optical control of topologically ordered phases. $tMoTe_2$ is a direct bandgap semiconductor with a strong exitonic response[29]. Due to the spin-valley coupling inherited from the monolayers[30], $tMoTe_2$ possesses valley dependent circularly polarized optical response. This enables helicity-selective optical access to the spin/valley degrees of freedom and the associated spontaneous ferromagnetism which arises from strong electron-electron interactions in the flat band[29,31]. The FQAH properties of $tMoTe_2$, such as the sign of Chern number and the chirality of the edge states, are determined by the ferromagnetic polarization. Thus, control of ferromagnetism would facilitate simultaneous control of the FQAH effects. Such a control of QAH effect has been realized by scanning probe manipulation of ferromagnetic domains in magnetically doped topological insulator thin films[32]. In $tMoTe_2$, prior work has demonstrated the strong coupling between the helicity of light, exitonic response, and QAH states[31,33], showing the promise of optical control of strongly correlated topology in this system.

Here, we demonstrate helicity-selective optical training, direct switching, and dynamic modulation of the moiré Chern ferromagnetism, as well as optical writing of CI/FCI domains in $tMoTe_2$. We fabricated dual gated $tMoTe_2$ using graphite as an optically transparent top gate (see Methods). A circularly polarized laser in resonance with trion states is employed as a pump to control the ferromagnetic states. We then perform helicity resolved photoluminescence (PL) measurements, using linearly polarized above-bandgap HeNe laser excitation as a probe of the ferromagnetism (Fig. 1a). As established in prior reports, in the ferromagnetic phase, the trion PL in $tMoTe_2$ is perfectly circularly polarized, with its helicity determined by the ferromagnetic polarization[31,33]. Two devices were used in our investigation. Device 1 with a twist angle of 3.7° is the same device measured in the initial report of FQAH effect[17]. Device 2 with a twist angle of 3.5° was fabricated from recently developed high quality $MoTe_2$ crystal[34]. Measurements are performed at 1.6 K, unless otherwise specified. The optical excitation power is measured before entering the cryostat, which is about three times that arriving at the sample.

## Optical training of Chern magnetism

We first present the deterministic preparation of ferromagnetic polarization by optical training. Data is from Device 1, with its basic properties shown in Figs. 1b&c. Fig. 1b is the reflective magnetic circular dichroism (RMCD) measurements versus filling factor $v$ and electric field $D/\varepsilon_o$. Here, $v$ represents the number of carriers per moiré unit cell. The data is taken at zero magnetic field (i.e. $B = 0$) after initializing the ferromagnetic polarization with a magnetic field $B$ (e.g. 0.5T). The phase space of non-vanishing RMCD signal highlights the spontaneous ferromagnetic order and its gate tunability. The results also show that once the ferromagnetic polarization is initialized by an external magnetic field, sweeping doping and electric field in and out of the ferromagnetic states does not change the ferromagnetic orientation. This observation, consistent with prior reports[16, 28], suggests that there is residual spin polarization - possibly defect spins - which provides a memory effect of the system. Figure 1c is the trion PL intensity plot versus photon energy and $v$. The quenched PL near $v$ of -3/5 and -2/3 correspond to the FCI states, while the same near $v = -1$ is the CI (or QAH) state.[17]

The optical training protocol is depicted in Fig. 1a. We start with the application of a small magnetic field (e.g. 0.5 T) to orient the spins in the same direction while in the ferromagnetic phase, and then switch off the field. After sweeping the gates to a value where the system becomes non-ferromagnetic (i.e. no spontaneous circularly polarized PL), a circularly polarized pump is turned on. We found that the choice of gate voltage (or filling factor $v$) for the non-ferromagnetic regime is not critical. For simplicity, we set $v = 0$ for the presented data in Figs. 1 and 2. In contrast, pump photon energy plays a critical role, as optical training is only effective near resonance with the trion (Extended Figure 1). The helicity of the light is chosen to inject spin/valley polarization opposite to that initialized by the magnetic field. While illumining tMoTe$_2$ by the pump light, we sweep the gates to a chosen value of $v$ where a Chern insulator state forms. Finally, the pump laser is turned off. We probe the ferromagnetic polarization of the prepared state using helicity resolved PL with a linearly-polarized low-power HeNe laser excitation (~ 1 nW) without perturbing the system. In essence, the procedure is similar to optical training of magnetization achieved by cooling a magnetic material through its Curie temperature[5]. The difference in our case is that rather than tuning the temperature, we tune the carrier density from non-ferromagnetic to ferromagnetic states, with a fixed temperature far below the Curie temperature.

Figures 1d illustrates the optical training results at $v = -1$, *i.e.* the QAH state. Before training, the ferromagnetic polarization was initialized by a positive magnetic field to produce trion PL with left circular polarization ($\sigma^-$). This corresponds to a +K valley polarized hole population[31]. The degree of circular polarization, $\rho = \frac{PL(\sigma^+) - PL(\sigma^-)}{PL(\sigma^+) + PL(\sigma^-)}$, is near unity (-1), as evidenced by dominant $\sigma^-$ polarized PL at zero magnetic field in the left panel of Fig. 1d. Here, $PL(\sigma^+)$ ($PL(\sigma^-)$) is the $\sigma^+$ ($\sigma^-$) polarized trion PL intensity. We then implement the optical training protocol, as described above, with left circularly polarized ($\sigma^-$) pumping. The pump power is 800 nW and photon energy is 1.11 eV. The optical training leads to the robust switching of the trion PL helicity from $\sigma^-$ to $\sigma^+$ with $\rho$ of about +1 (Fig. 1d right panel, and Extended Data Fig. 2). The same measurements are also performed on the -2/3 FCI state, as shown in Fig. 1e. The observed complete switching of trion PL helicity by optical training demonstrates the flipping of the ferromagnetic polarization, and thus the sign of the underlying Chern numbers for both CI and FCI states.

The optical training of ferromagnetism appears to be most effective at the CI/FCI states. Figure 2a shows $\sigma^+$ (left panel) and $\sigma^-$ (middle panel) polarized trion PL versus $v$, with extracted $\rho$ in the right panel. The optical training protocol is performed at each value of $v$. Evidently, only near $v = -1$ and -2/3, the helicity of the trion PL (and thus the underlying ferromagnetic polarization) is flipped compared to the magnetic field-initialized state. This observation can be understood by optical orientation of valley-polarized holes, which creates an optical torque. Figure 2b illustrates the creation of valley polarized trions by the $\sigma^-$ polarized resonant pumping, in which an electron-hole pair excited in the -K valley binds to an existing hole in the +K valley, driven by a repulsive intervalley exchange interaction[37]. Due to more efficient intervalley scattering of electrons than holes — enabled by the much smaller spin splitting in the conduction band than the valence band — a time-reversed partner trion can form. Its recombination depletes the initially existing holes in the +K valley. The microscopic mechanism of the electron scattering may be defect-assisted or phonon-assisted[36]. Consequently, $\sigma^-$ resonant excitation leads to a net gain in hole population in -K relative to +K valley, promoting the formation of a spin-down ferromagnetic domain (Fig. 2b). The valley population of holes depends on excitation power, trion lifetime, and intervalley electron/hole scattering rate. The trion lifetime likely increases at CI/FCI states due to an energy gap[38]. These factors align with our finding that optical training is most effective at the CI/FCI states.

The above interpretation is further supported by both electric field ($D/\varepsilon_o$) and power dependence. Figure 2c shows $\rho$ versus $D/\varepsilon_o$ at the QAH state ($v = -1$, See Extended Data Fig. 3 for polarization-resolved spectra). The data is taken with $\sigma^+$ polarized optical training, while the system was initialized at $B = -0.5$T (corresponding to $\sigma^+$ polarized trion PL). The results reveal that the helicity of the trion PL flips only within a moderate electric field range (-90mV/nm< $D/\varepsilon_o$ <40mV/nm), significantly narrower than the full ferromagnetic regime.[17,29] As explained in prior report, the Chern gap becomes smaller and eventually closes during the electric field-induced QAH to correlated insulator phase transition[15]. The reduction of the gap accompanies a decrease of trion radiative lifetime, which reduces the optical torque from the training. Similar electric field dependent optical training is observed for the -2/3 FCI state (Fig. 2d, Extended Data Fig. 3). The inset of Fig. 2c further shows dependence of $\rho$ on the optical training power at the QAH state. The training is only effective above a threshold, as only sufficiently large spin/valley polarization generated by optical pumping can polarize the ferromagnetic domain.

One would expect that if the excitation power were large enough, it may produce sufficient optical torque to overcome the energy barrier between two ferromagnetic polarizations and thus directly switch between them without the need to sweeping the gates. This is indeed the case, as shown in Extended Data Fig. 4. For the rest of the paper, we focus on direct optical switching of ferromagnetic polarization in Device 2 (see basic properties in Extended Data Fig. 5). Device 2 is fabricated from new crystals with nearly two orders of magnitude lower defect density than those from which Device 1 was fabricated. Recently, we have demonstrated dissipationless FCI with quantized $R_{xy}$ and vanishing $R_{xx}$ using devices made from these new crystals[34]. We found that this type of higher-quality device enables optical switching of ferromagnetism with a substantially lower power threshold than Device 1. The differences between Devices 1 and 2 are presumably due to defect density-dependent spin/valley relaxation rates, which are reduced in Device 2 because of the drastically improved crystal quality.

## Switching Chern magnetism by direct optical pumping

Figure 3a depicts the experimental procedure. At ferromagnetic state of fixed $v$, we employed circularly polarized optical pumping to inject spin/valley polarization. The pump is then turned off, and the ferromagnetic polarization is subsequently read out by a weak linearly polarized optical excitation (1 nW HeNe laser) with helicity resolved PL detection. Figure 3b compares the circular polarization-resolved trion PL before (left) and after (right) the direct optical pumping at the QAH state (*i.e.* $v = -1$). The pump is near resonance with the trion state with a power of 2 µW. The opposite helicity of trion PL before and after pump demonstrates a successful direct optical switching of ferromagnetic polarization, and thus the Chern number in the QAH state. The degree of circular polarization both before and after the pumping reaches unity. This implies the polarization switching is uniform under the beam spot without appreciable magnetization domain effects. Domain physics will be further discussed in the next section.

We investigate the direct optical switching of ferromagnetic polarization versus $v$. For simplicity, we present the case where ferromagnetic states are initialized to emit $\sigma^-$ polarized trion PL. Figure 3c shows $\rho$ versus $v$ and photon energy at zero magnetic field after $\sigma^-$ polarized optical pumping at each $v$ (Extended Data Fig. 6). An extracted line trace is also presented in Fig. 3d. The data show that for the phase space near $v = -1$ and $-2/3$, the trion PL helicity is successfully switched from $\sigma^-$ to $\sigma^+$. This demonstrates the direct optical switching of ferromagnetic polarization. Notably, the trion PL helicity either reduces or is not flipped for $v$ near -0.8 and below -1/2. Transport measurements have revealed gapless behavior for these fillings[15]. The state near $v=-0.8$ was attributed to ferromagnetic metal (or anomalous Hall metal)[15,33]. The state near -1/2 behaves consistently with a putative composite Fermi liquid state[15,31,39,40]. The absence of the energy gap of these states facilitates the radiative decay of the valley-polarized trions, which as a result do not provide enough optical torque to fully flip magnetic domains. The observation here is consistent with the optical training results presented in Fig. 2a.

The direct optical switching of ferromagnetic polarization exhibits similar pump power (Fig. 3e) and electric field dependence (Fig. 3f) as the optical training measurement discussed previously. As shown in Fig. 3e, direct optical switching of full polarization occurs for excitation powers above ~600 nW for 1.117 eV excitation. Note that the threshold of the excitation power is photon-excitation energy dependent, since the optical absorption is highly sensitive to detuning from the trion resonance. Figure 3f shows $\rho$ versus $D/\varepsilon_o$ for the QAH state near $v$ of -1 and the -2/3 FCI state (Extended Data Fig. 6). Measurements of $\rho$ both before and after optical pumping are shown for comparison. It is evident that the electric field range of effective optical switching of ferromagnetic polarization is slightly smaller than the range for the QAH phases. This observation is consistent with those in Figs. 2c&d resulting from optical training. It further supports our interpretation that an energy gap increase the lifetime of valley-polarized trion resulting from circularly polarized optical pumping.

The observation of direct optical switching of ferromagnetism in FCI is quite remarkable. Flipping ferromagnetic polarization by light is a known challenge in the community. Notable results include those utilizing intense ultrafast optical pulses to switch metallic magnets near the critical temperature of magnetic order[41-44]. In those cases, photoexcitation-induced heating plays a key role. In our experiment, the temperature of 1.6 K is much smaller than Curie temperature of 14 K for $v$ near -1 and 4.5 K for $v$ near -2/3. In addition, the optical pumping power up to a few µW produces negligible heating effect on the sample. Thus, the switching mechanism in our measurements is distinct from those relying on photo-heating effects. As stated in Fig. 2, the ferromagnetic polarization switching is due to the effective optical torque induced by valley-selective hole injections from circularly polarized resonant pumping. Besides the assistance of strong excitonic response and gap protected spin/valley polarization, which result in large torque, the moiré superlattice also reduces the spin density (~1 electron spin per moiré unit cell), which is at least two orders of magnitude lower than in atomic ferromagnets. This lowers the threshold of the critical torque and pump power.

**Modulation and patterning of (F)QAH domains**

The direct optical switching of ferromagnetic polarization enables dynamic manipulation of the underlying Chern insulator states. We modulate the helicity of the optical pump between left and right circular polarization. During the helicity switching, we block the pump and probe the ferromagnetic polarization by measuring the spontaneous helicity of the trion PL. Figure 4a shows the result for $v = -2/3$ FCI state. The left panel is $\sigma^+$ polarized PL; the mid-panel is $\sigma^-$ polarized PL. The resulting $\rho$ is plotted in the right panel, with its line trace versus time shown in Fig. 4b. The observed periodic switching of the PL helicity demonstrates the successful dynamic optical manipulation of ferromagnetic polarization. Similar results are also obtained for the $v = -1$ QAH state (see Extended Data Fig. 7). The helicity modulation in our experiment is performed on the time scale of minutes, demonstrating the potential of tMoTe$_2$ as optically controllable topological magnetic memory[4]. An interesting future direction is to explore ultrafast manipulation of the ferromagnetism and correlated topological phases by circularly-polarized optical pump pulses.

We also performed spatial mapping of $\rho$ following direct optical switching, enabling visualization of the resulting ferromagnetic domains. Figure 4c presents maps of $\rho$ of the -2/3 FCI state under varying excitation powers (see Extended Data Fig. 8 for full data set). At low power (400 nW) excitation, the domain size is approximately 2 μm, comparable to the optical beam spot size. As the excitation power increases, the domain expands in size, demonstrating optical writing and control of a ferromagnetic and thus a FQAH domain. Similar domain writing behavior in the QAH state is shown in Extended Data Fig. 9. Note that the mapping was conducted over a 20-hour period, underscoring the robustness of the optically written domains. Since edge currents are expected at the domain boundaries[32], light-coupled scanning probe techniques, such as exciton resonant microwave impedance microscopy, offer a promising route to directly probe these optically induced QAH/FQAH domain[19]. With the integration of advanced spatial light modulation, our results point towards the possibility of preparing on-demand Chern insulator domains via optical writing.

In this work, we successfully translate the light helicity to the spin/valley-polarized holes, enabling optical control over the QAH/FQAH states. We emphasize that this is a control of the many-body states with energy gaps in the order of meV by using photons at the energy of about 1 eV. Such optical control usually requires the photon energy to be resonant with the many-body gap. In tMoTe$_2$, the unique coupling between spin, valley, charge, and photon enables the control of the low energy many-body states with three orders of magnitude difference in photon and gap energy. Our work constitutes a foundation for optical creation of valley-polarized states with anyonic quasiparticle excitations – an essential ingredient for realizing anyonic quantum memory and computation[11].

## Methods

### Device fabrication

The tMoTe$_2$ Device 1 (3.7º) used is the same device used in Ref. 17. It was fabrication from commercially available bulk crystal and the fabrication procedure was discussed in Ref. 17 in detail. Device 2 (3.5º) was fabricated from MoTe$_2$ crystal grown in house. Device assembly began with the preparation of the back gate by sequentially picking up a bottom hBN dielectric, a bottom gate graphite electrode and then dropping down onto the substrate at ~170 °C using PC dry-transfer techniques. The heterostructure was then immersed in chloroform for 10 minutes to dissolve the polymer, followed by AFM cleaning to remove residual PC. Next, a thin hBN flake, a strip of contact graphite, and half of a monolayer MoTe$_2$ flake were sequentially picked up. The second half of the MoTe$_2$ monolayer was then rotated to the target twist angle before being incorporated, forming the moiré heterostructure, which was subsequently melted onto the prepared back gate. As MoTe$_2$ is air sensitive, its exfoliation and the tMoTe$_2$ assembly were both carried out in an argon filled glovebox. To ensure a clean interface, we use AFM cleaning to squeeze out trapped air bubbles. Finally, top gate graphite and top hBN dielectric are transferred on top of the whole stack before patterning gold wire bonding pads using standard electron beam lithography, followed by E-beam evaporation to complete the device fabrication.

### Optical measurements

All measurements were performed in a closed-loop magneto-optical cryostat (attoDRY2100) with a base temperature of 1.6K. A high-NA nonmagnetic cryogenic objective was used to focus light on and collect light from the sample. Unless otherwise stated, all measurements were done at 1.6 K. Optical control measurements were performed using a pump beam in resonance with the trion energy to initialize the desired magnetization direction. The pump was then switched off, and a linear polarized probe beam (632.8 nm HeNe laser) was subsequently used to measure circular polarization-resolved PL to read out the magnetization direction. For the pump beam, a supercontinuum source (NKT SuperK Fianium FIU-15) was used. The light source was filtered with a home-built single grating double subtractive monochromator to the desired wavelength with a linewidth less-than 1 nm. The pump passes through a linear polarizer and a quarter-waveplate to be either $\sigma^+$ or $\sigma^-$ polarized before being focused onto the sample with a roughly 2 µm beam spot. For the optical training of ferromagnetism, the pump beam was on while the gate voltages were swept from 0 V to their targeted values. The pump was then switched off before PL was used to probe ferromagnetic polarization. For the direct optical switching measurements, the gate voltage was fixed at targeted value, and the pump beam was unblocked for about 2 seconds. The pump beam was then blocked, and the ferromagnetic polarization was read out by helicity resolved PL excited by a linearly polarized 632.8 nm HeNe laser. The probe beam was focused onto the sample with an approximately 1 µm beam spot. Photoluminescence from the sample passed a quarter-waveplate and a linear polarizer to select either $\sigma^+$ or $\sigma^-$ polarized emission. The signal was passed through a spatial filter with a 75 µm pinhole and then enters a spectrometer (Teledyne-Princeton Instruments SpectraPro HRS-500), where it was dispersed by a diffraction grating (300 groves/mm, 1 µm blaze or 600 groves/mm, 1 µm blaze) and measured by a liquid-nitrogen-cooled infrared InGaAs photodiode array (PyLoN-IR 1.7).

RMCD measurements were performed on resonance with the trion energy using the same light source described above for the pump beam. The filtered light was chopped at $f_c$ = 1 kHz and its polarization modulated between left- and right-circularly polarized at $f_p$ = 50 kHz by a photoelastic modulator. The light was then focused onto the sample with an approximately 2 μm beam spot. The reflected light was collected and sent to an InGaAs avalanche photodiode, which had its output demodulated by two lock-in amplifiers to obtain the signals $I_c$ and $I_p$ at the frequencies $f_c$ = 1 kHz and $f_p$ = 50 kHz, respectively. The RMCD signal was given by $\Delta R/R = I_c/(J_1(\pi/2) \times I_p)$, where $J_1$ is the first-order Bessel function.

**Acknowledgements:** The authors thank Atac Imamoglu and Tomasz Smolenski for the insightful discussion. This project is mainly supported by the U.S. Department of Energy (DOE), Office of Science, Basic Energy Sciences (BES), under the award DE-SC0012509. The fabrication and measurement of the 3.5° device is partially supported by Vannevar Bush Faculty Fellowship (Award number N000142512047). Bulk MoTe$_2$ crystal growth and characterization for the 3.5° device is supported by Programmable Quantum Materials, an Energy Frontier Research Center funded by DOE BES under award DE-SC0019443. T.C. acknowledges the support by the U.S. Department of Energy, Office of Basic Energy Sciences, under Contract No. DE-SC0025327 for part of the theoretical analysis. The authors also acknowledge the use of the facilities and instrumentation supported by NSF MRSEC DMR-2308979. K.W. and T.T. acknowledge support from the JSPS KAKENHI (Grant Numbers 21H05233 and 23H02052) , the CREST (JPMJCR24A5), JST and World Premier International Research Center Initiative (WPI), MEXT, Japan. XX acknowledges support from the State of Washington funded Clean Energy Institute and from the Boeing Distinguished Professorship in Physics.

**Author contributions:** XX conceived and supervised the experiment. JC fabricated the 3.7° devices. WL fabricated the 3.5° degree device. WH and WL performed the measurements with assistance from EA and HP. WH, WL, DX, TC, XX analyzed and interpreted the results. TT and KW synthesized the hBN crystals. CH and JHC grew and characterized the bulk MoTe$_2$ crystals, which were used for the fabrication of 3.5° devices. XX, WL, WH, EA, TC, JC, and HP wrote the paper with input from all authors. All authors discussed the results.

**Competing interests:** The authors declare no competing interests.

**Data availability:** Source data that reproduces the plots are provided with this paper. All supporting data for this paper and other findings of this study are available from the corresponding author upon reasonable request.

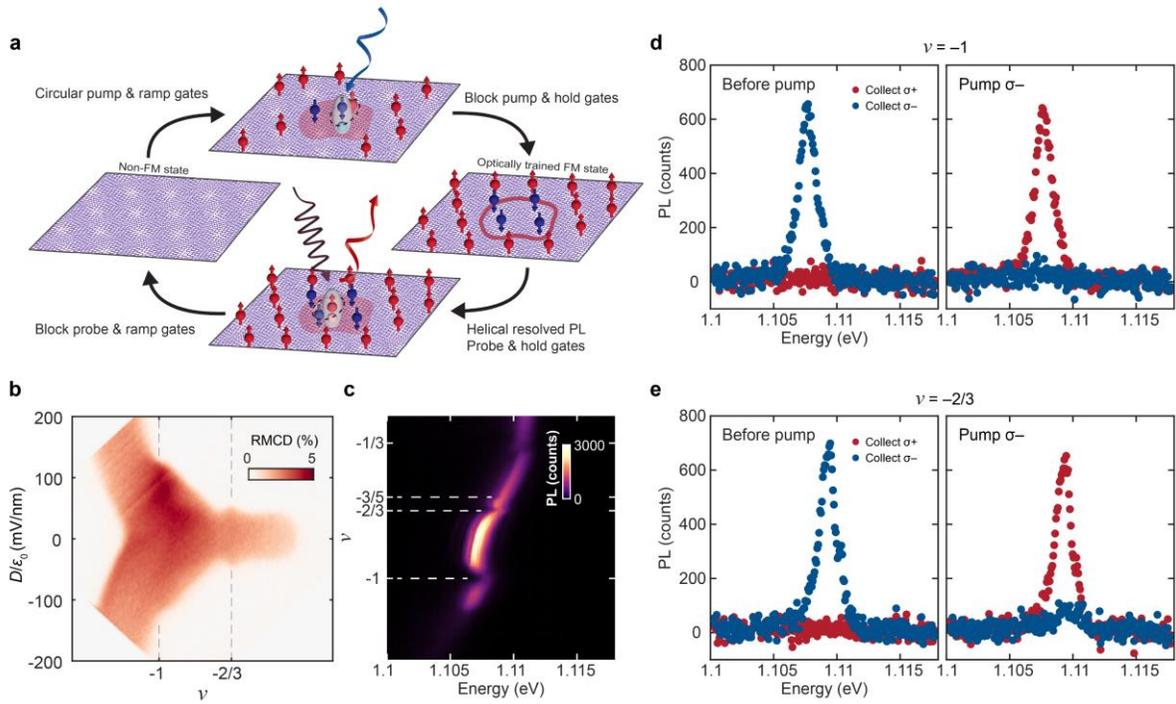

**Fig. 1 | Optical training of moiré Chern ferromagnetism. a,** Schematic of the experiments and the flow diagram of the optical control and measurements. See text for details. **b**, Reflective magnetic circular dichroism (RMCD) signal versus electric field $D/\varepsilon_o$ and filling factor $v$ at zero magnetic field (B=0). **c**, Trion photoluminescence intensity versus $v$ and photon energy, with marked integer and fractional Chern insulator states. **d**, Left panel: circular polarization resolved trion PL spectra at filling factor $v$ = -1, *i.e.* the quantum anomalous Hall state. The state, initialized by a small positive magnetic field, has spontaneously $\sigma^-$ polarized PL. Right panel: circular polarization resolved trion PL after $\sigma^-$ optical training, following the protocol described in (a). The switching of PL helicity demonstrates the flipping of ferromagnetic polarization by optical training. **e**, The same measurements as (d), but at the $v$ = -2/3 FCI state. The PL in (d&e) are excited with linearly polarized HeNe laser.

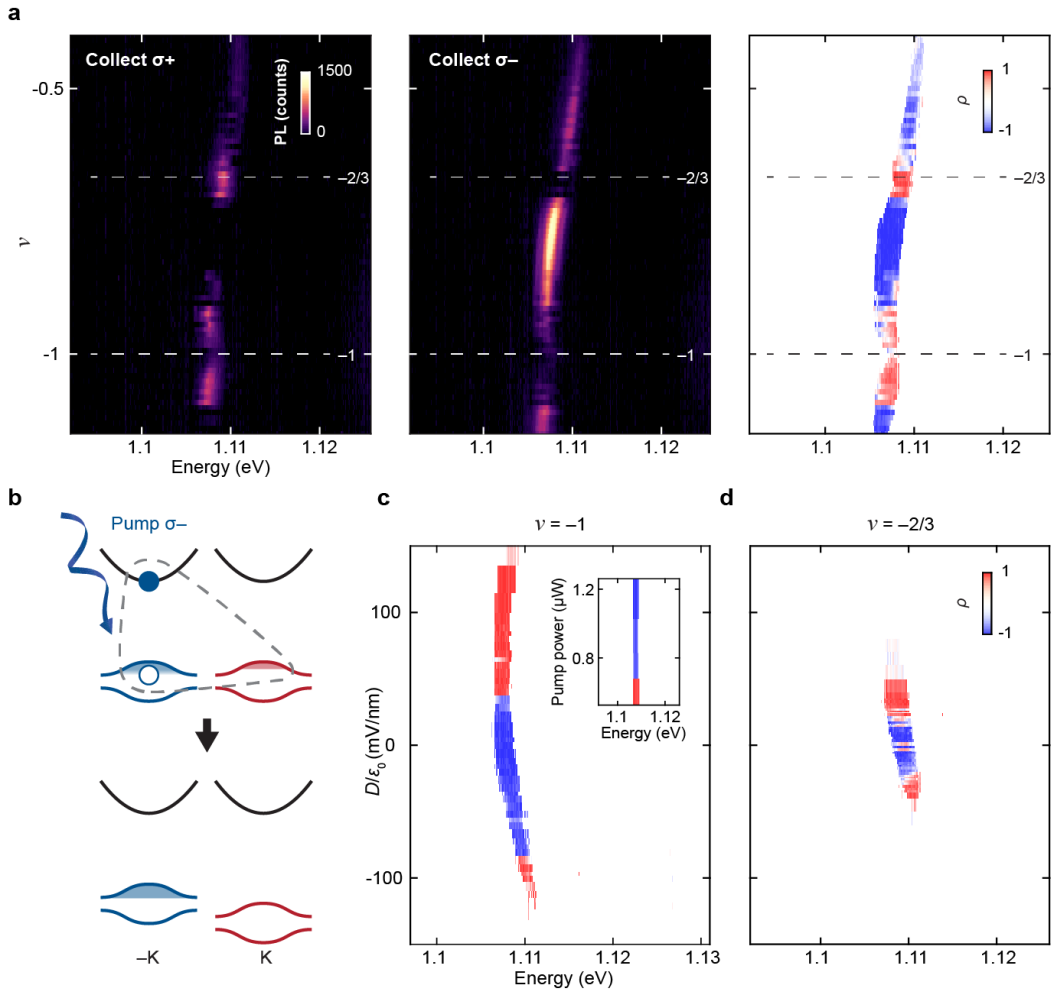

**Fig. 2 | Doping, pump power, and electric field dependent optical control. a**, Trion PL intensity plot with $\sigma^+$ (left panel) and $\sigma^-$ (middle panel) polarized detection. The data were taken with system initialized with +0.5 T magnetic field before employing the optical pumping protocol with $\sigma^-$ polarized light. The resulting degree of circular polarization ($\rho$) is shown in the right panel. It reveals that optical training of ferromagnetic polarization is only effective at the integer and fractional Chern insulator states. **b**, Schematic of optical orientation of spin/valley polarization by resonant optical pumping of trion states. While sweeping the gate to the incompressible states, due to the gap enhanced trion lifetime and electron-hole recombination in the opposite valley, the optically injected holes function as optical torque and thus determine the polarization of valley ferromagnetism. **c**, Degree of circular polarization $\rho$ versus $D$ and photon energy after optical training at $v=-1$. The training protocol is employed with -0.5T magnetic field initialization, *i.e.* the system begins with spontaneous magnetization favoring $\sigma^+$ trion luminescence, opposite to the data in (a). Inset: $\rho$ versus $\sigma^+$ polarized optical training power. The helicity switches from $\sigma^+$ to $\sigma^-$ near 700 nW, which photo-injects enough spin/valley polarization to train the ferromagnetic polarization. **d**, The same measurement as (c) but at the -2/3 FCI state. Both (c) and (d) show that the optical training is only effective at small electric field, where the Chern insulator gap is large. At large electric field, Chern gap reduces and thus the optical training becomes ineffective.

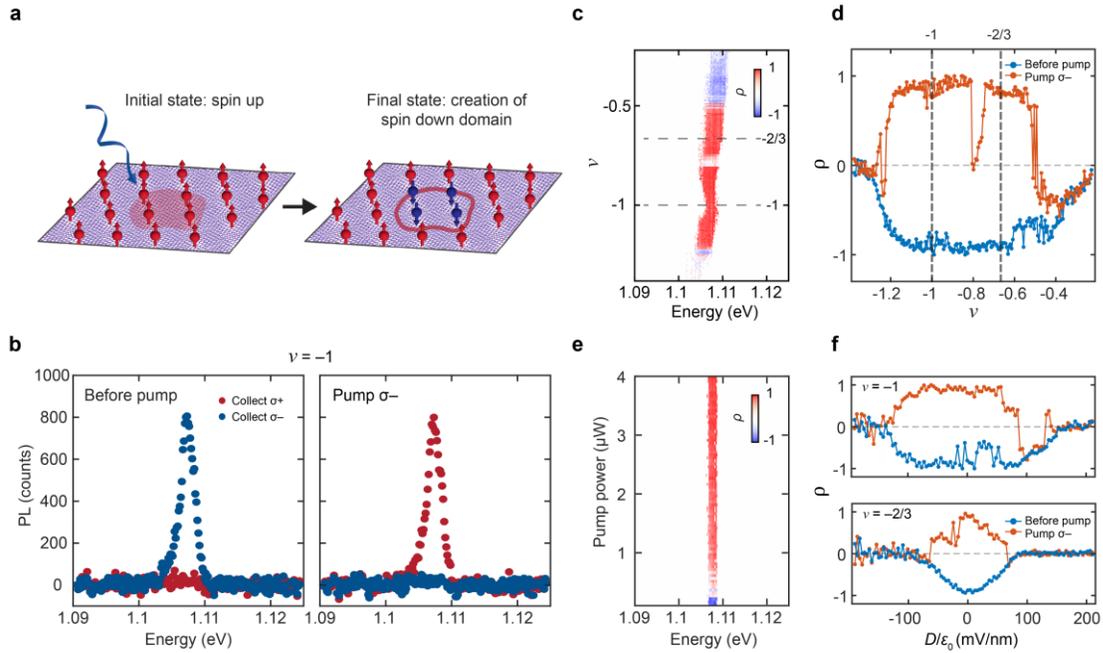

**Fig. 3 | Direct optical switching of moiré Chern ferromagnetism.** All data are taken at B=0. **a**, Experimental scheme of chiral optical pumping. Left panel: circularly polarized pump directly flips the ferromagnetic moment of the Chern insulator states. Right panel: probing the ferromagnetic polarization by helicity resolved photoluminescence with linearly polarized excitation. **b**, Helicity resolved PL spectra before (left) and after (right) the application of optical pumping. The data are taken near the $v$=-1 quantum anomalous Hall state. The switching of trion PL helicity demonstrates the successful direct optical switching of moiré Chern ferromagnetism. **c**, $\rho$ versus $v$ and photon energy after employing circularly-polarized optical pumping. Before pumping, the trion PL favors $\sigma^-$ polarization. **d**, The trion helicity $\rho$ versus $v$ extracted from (c). $\rho$ before optical pumping is plotted for comparison (blue curve). Both (c) and (d) show that optical switching of ferromagnetism is most effective near the integer and fractional Chern insulator states, where the energy gap increases trion lifetime and preserves optically injected hole valley polarization. **e**, $\rho$ versus optical pumping power. Helicity switching occurs above ~600 nW. **f**, $\rho$ versus electric field for $v$ near -1 (top panel) and -2/3 (bottom) after the application of optical pumping. For comparison, $\rho$ before optical pumping is also shown.

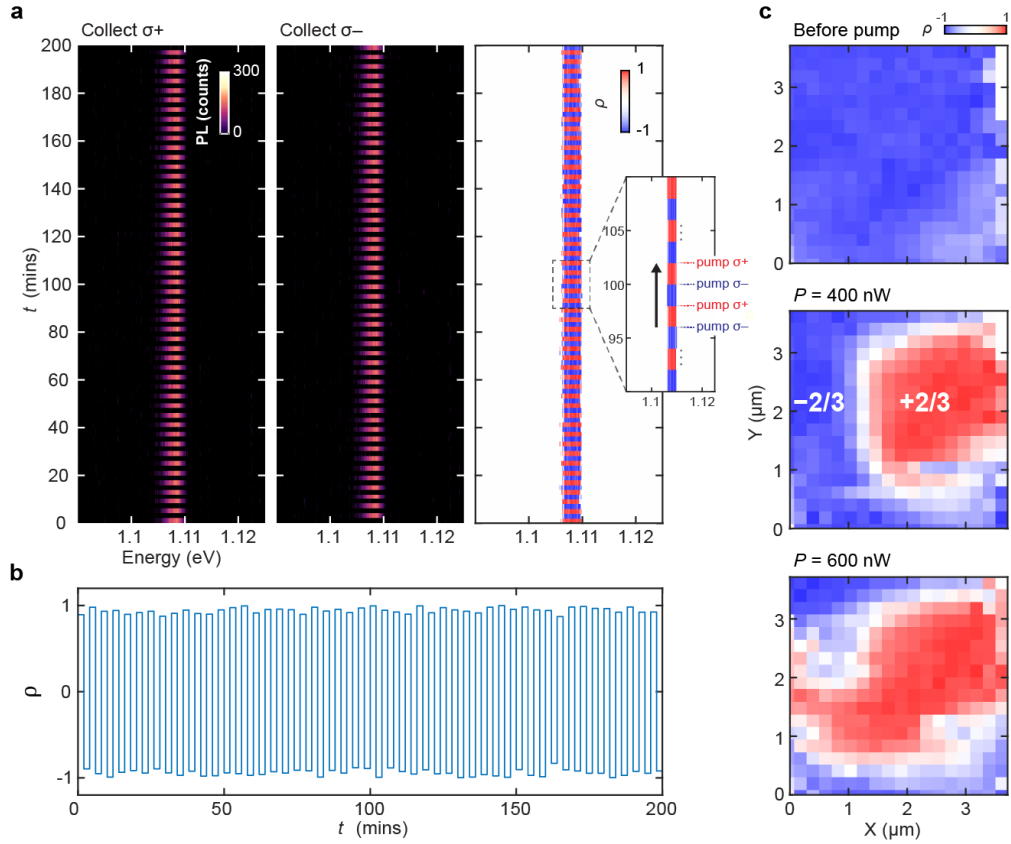

**Fig. 4 | Dynamic optical manipulation of the -2/3 FCI state. a,** $\sigma^+$ (left panel) and $\sigma^-$ (middle panel) resolved PL measurements as the helicity of optical pump is dynamically modulated between $\sigma^-$ and $\sigma^+$. During each helicity resolved PL detection, the pump is off. The right panel is $\rho$ extracted from the two left panels. The inset is the zoom-in plot of $\rho$ as pump helicity is dynamically modulated. **b,** Extracted $\rho$ versus time, demonstrating dynamic switching of FQAH ferromagnetic polarization. **c,** Spatial map of $\rho$ after circularly polarized optical pumping, revealing the optical writing of ferromagnetic and thus FCI domain (red color) with an opposite sign of Chern number (+2/3) compared to the state (-2/3) without pumping. The top, middle, and bottom panels are taken without and with optical pumping at 400 and 600 nW, respectively.

# Extended Data for

# Optical Control of Integer and Fractional Chern Insulators

**Authors:** William Holtzmann[1*], Weijie Li[1*], Eric Anderson[1], Jiaqi Cai[1], Heonjoon Park[1], Chaowei Hu[1], Takashi Taniguchi[2], Kenji Watanabe[3], Jiun-Haw Chu[1], Di Xiao[4,1], Ting Cao[4], and Xiaodong Xu[1,4#]

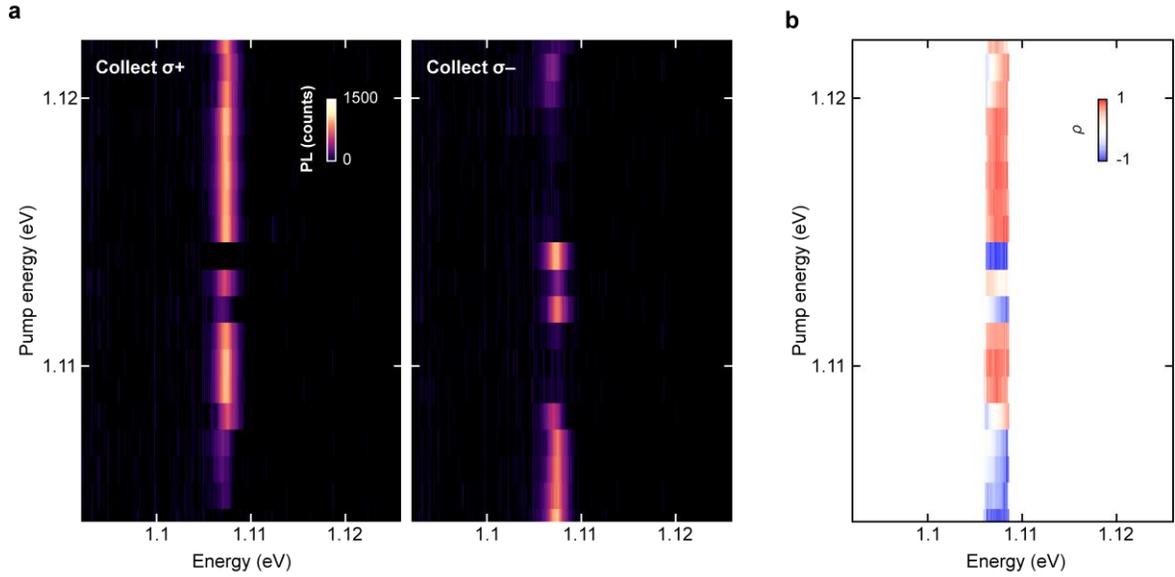

**Extended Data Fig. 1 | Photon energy dependent optical training effect. a**, $\sigma^+$ (left panel) and $\sigma^-$ (right panel) polarized PL versus pump energy at the quantum anomalous Hall state ($v$=-1). The system is first initialized to a ferromagnetic state which emits $\sigma^-$ polarized light, and the measurement is taken after $\sigma^-$ polarized optical training for each photon energy. **b**, Degree of circular polarization ρ extracted from the data in (a). $\sigma^+$ polarized emission (i.e. red color) shows the photo-excitation energy which induces magnetization flip after optical training.

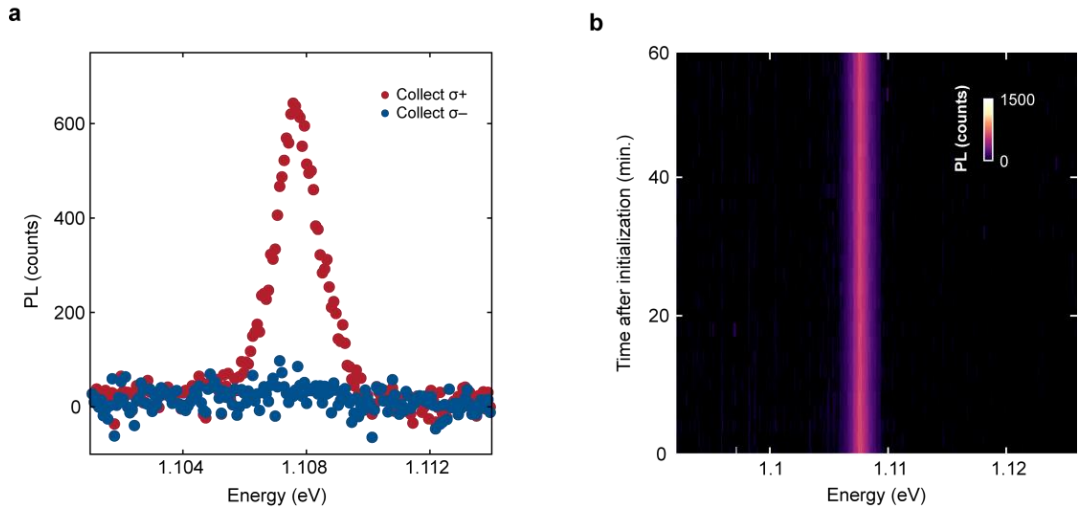

**Extended Data Fig. 2 | Robust ferromagnetism after optical training. a,** Circular polarization resolved PL spectra at $v$=-1 (quantum anomalous Hall state) after $\sigma^-$ polarized optical training. The data is taken at zero magnetic field. The PL emission is perfectly right circularly polarized. **b,** $\sigma^+$ polarized PL as a function of time after optical training. The PL spectra show little change over the measurement time frame of an hour, demonstrating the robust ferromagnetism after optical training.

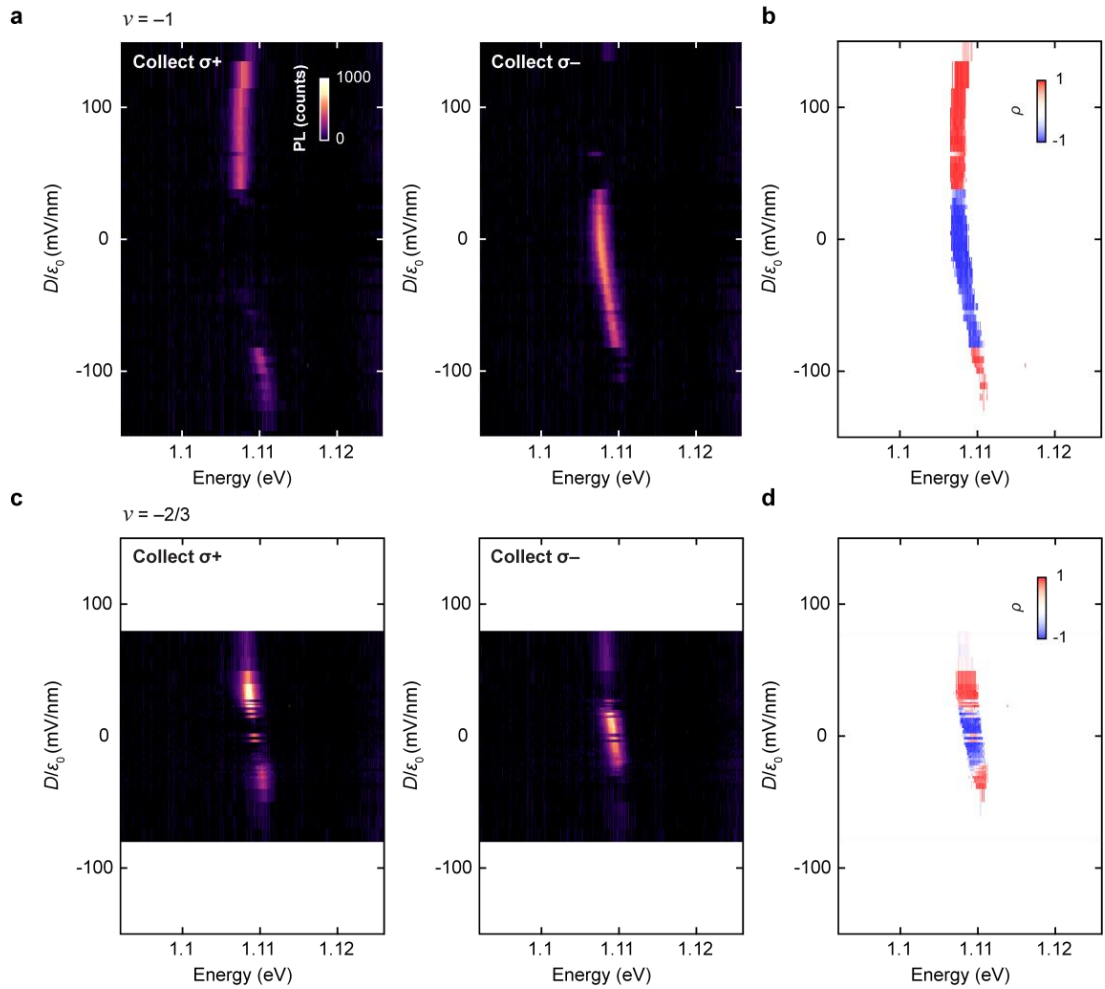

**Extended Data Fig. 3 | Electric field dependence of optical training effect. a,** $\sigma^+$ (left panel) and $\sigma^-$ (right panel) polarized PL versus electric field and photon energy in the quantum anomalous Hall state ($\nu$=-1). The system was initialized by -0.5T, resulting in $\sigma^+$ polarized PL emission before training. Then the data are taken after $\sigma^+$ polarized optical pumping. **b,** The resulting degree of circular polarization $\rho$ (the same plot as in Fig. 2c). The ferromagnetic polarization is flipped within a range of electric field (the blue color), which is smaller than the range for the existence of quantum anomalous Hall state. **c,** The same measurements as in panel (a) but for the -2/3 FCI state. **d,** $\rho$ versus electric field and photon energy for the -2/3 state (the same plot as in Fig. 2d).

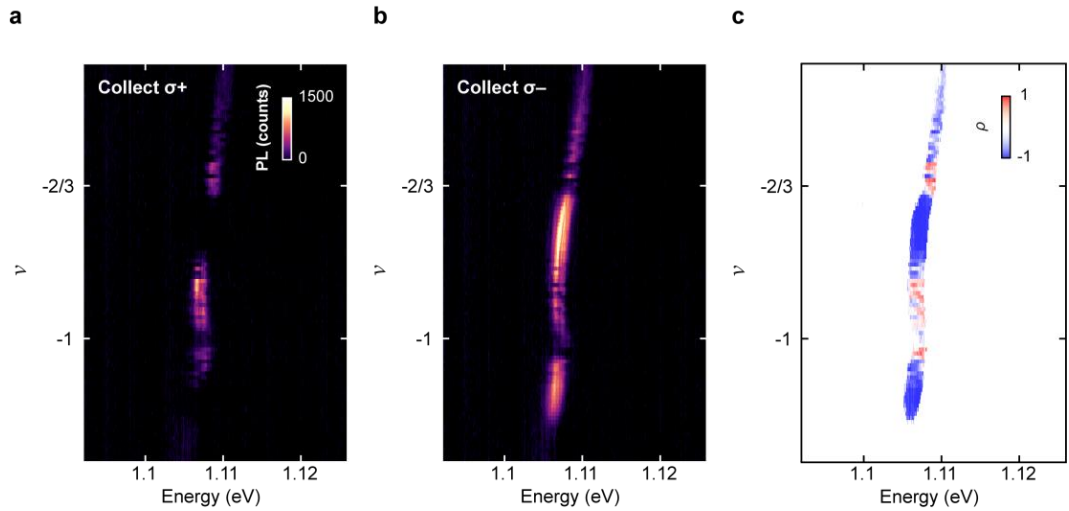

**Extended Data Fig. 4 | Direct optical switching of ferromagnetism of Device 1. a,b**, $\sigma^+$ (a) and $\sigma^-$ (b) polarized PL versus doping and photon energy at zero effective displacement field. The data is taken after $\sigma^-$ polarized optical pump at a power of 5 µW. **c**, The resulting degree of circular polarization $\rho$ for the data in (a&b). Weak direct optical switching of ferromagnetism is observed near -1 and -2/3 fillings.

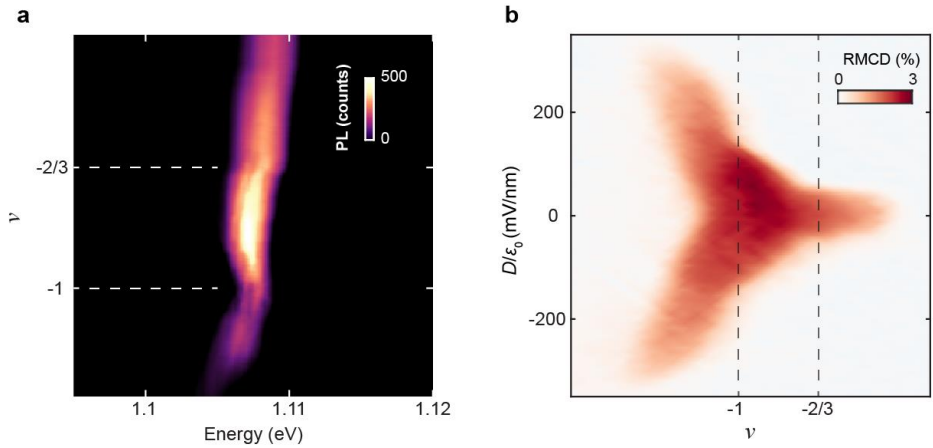

**Extended Data Fig. 5 | Photoluminescence and RMCD measurements of Device 2. a**, Trion PL intensity versus doping and photon energy. PL intensity is quenched at integer ($v=-1$) and fractional ($v=-2/3$) Chern insulator states. **b**, Reflective magnetic circular dichroism (RMCD) signal versus electric field $D/\varepsilon_o$ and filling factor $v$. The phase space with non-vanishing RMCD signal represents the spontaneous ferromagnetic phase, consistent with prior reports. All data are taken at a temperature of 1.6K and zero magnetic field. All data in the following Extended Data Figures are taken from Device 2.

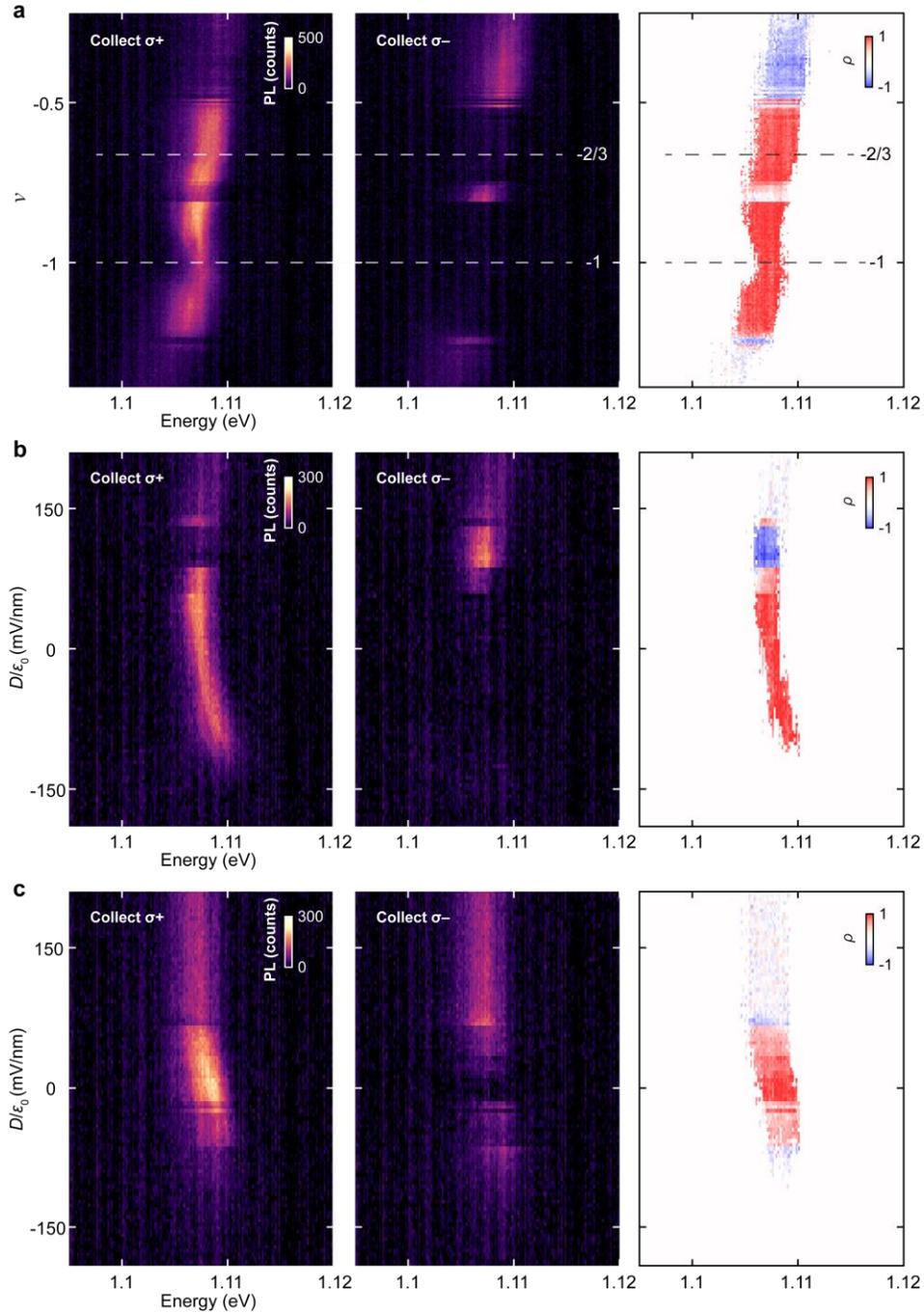

**Extended Data Fig. 6 | Helicity resolved PL versus doping after direct optical switching of ferromagnetism. a** Trion PL intensity plot with $\sigma^+$ (left panel) and $\sigma^-$ (middle panel) polarized detection, from which the plot in Fig. 3c of the main text is extracted (reproduced here at the right). **b**, $\sigma^+$ (left panel) and $\sigma^-$ (middle panel) polarized PL versus electric field and photon energy at $\nu$=-1. The resulting degree of circular polarization ρ is shown on the right. **c**, The same measurements as in panel (b) but for the -2/3 FCI state.

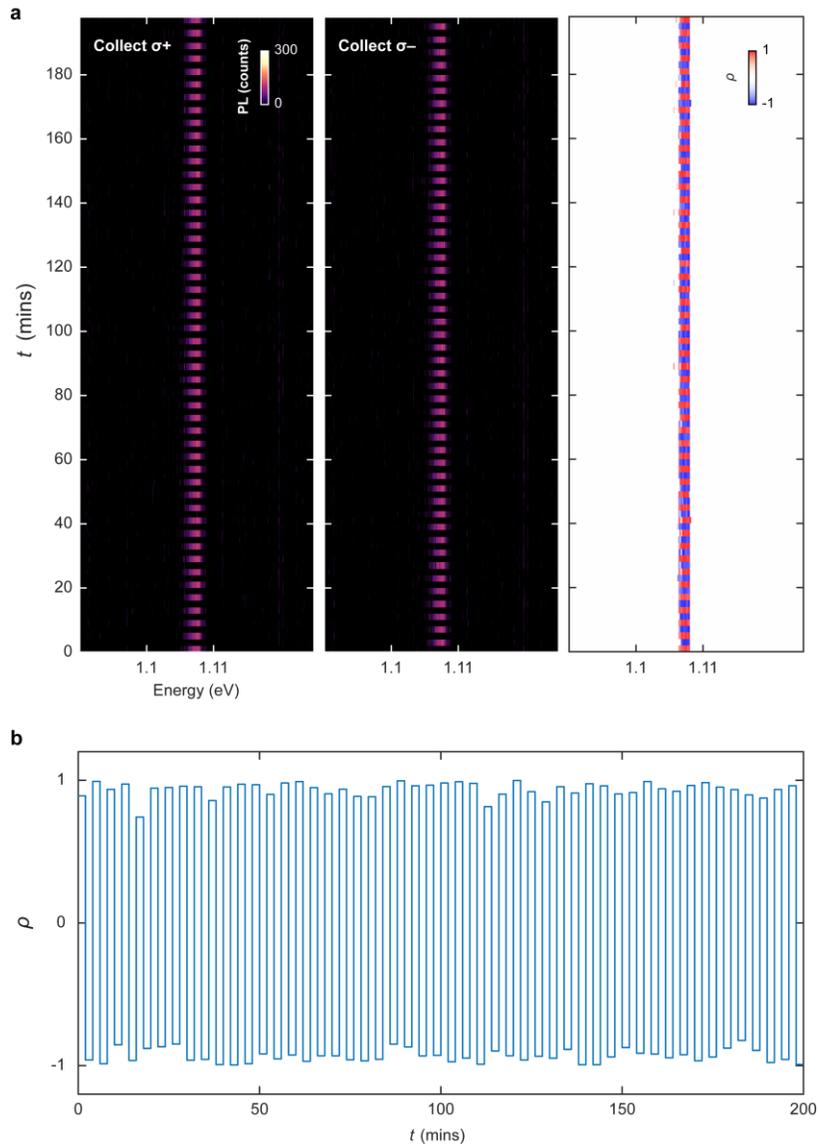

**Extended Data Fig. 7 | Dynamic optical switching of ferromagnetism at *v*=-1 quantum anomalous Hall state. a,** $\sigma^+$ (left panel) and $\sigma^-$ (middle panel) resolved PL measurements as the helicity of optical pump is dynamically modulated between $\sigma^-$ and $\sigma^+$. During each PL detection, the circularly polarized pump is off. The right panel is the degree of polarization $\rho$ extracted from the two left panels. **b,** Extracted $\rho$ vs time, demonstrating dynamic switching of ferromagnetic polarization of the QAH state.

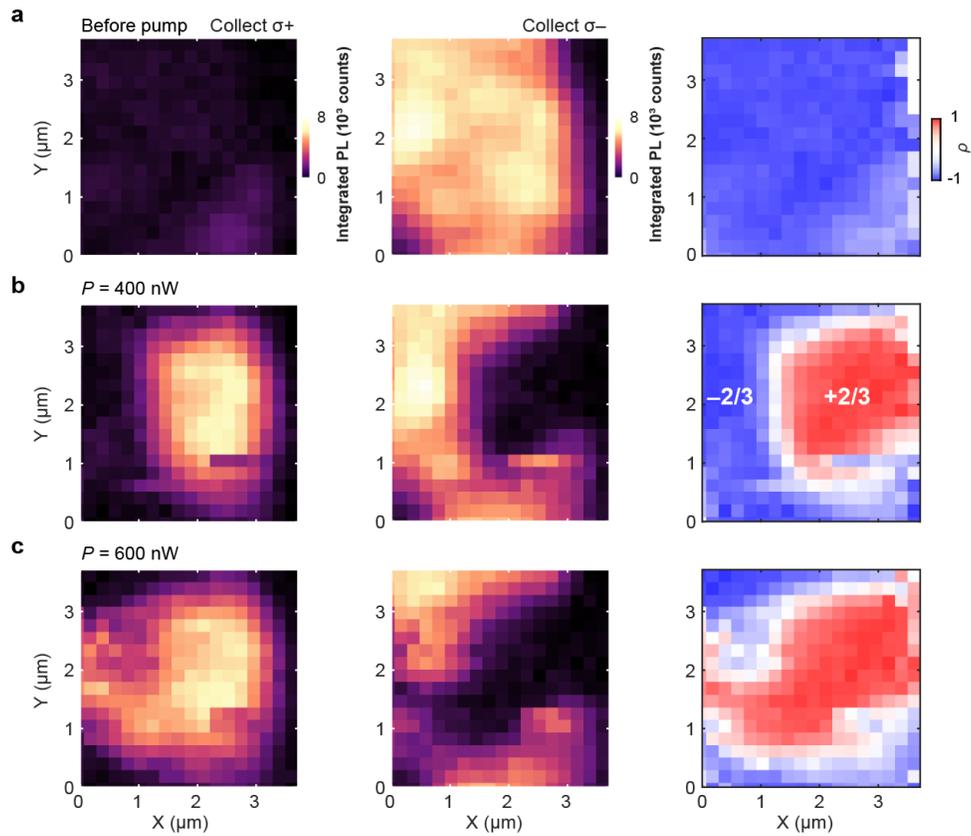

**Extended Data Fig. 8 | Spatially resolved FQAH domains created by direct optical pumping.** Spatially resolved PL with $\sigma^+$ (left column) and $\sigma^-$ (middle column) polarized detection at $v$=-2/3, which leads to the degree of circular polarization maps in the right column (also Fig. 4c in the main text). **a**, **b**, and **c**, are taken without and with $\sigma^-$ polarized optical pumping at 400 and 600 nW, respectively. The pump energy is 1.117 eV. The Chern numbers of the domains are indicated in the right column of (b).

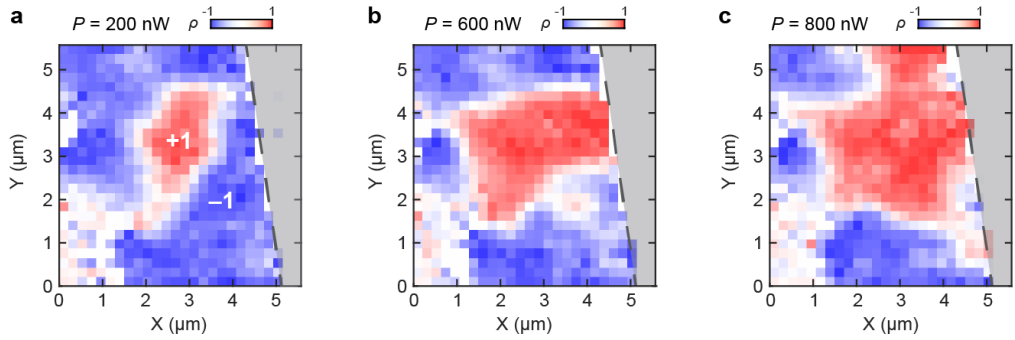

**Extended Data Fig. 9 | Spatially resolved QAH domains created by direct optical pumping.** Spatial map of ρ after circularly polarized optical pumping at ν=-1, revealing the ferromagnetic and thus QAH domain (red color) by optical writing. Maps are taken after σ⁻ polarized optical pumping at **a**, 200 nW, **b**, 600 nW, and **c**, 800 nW, respectively. The pump energy is set to be 1.11 eV. Dashed lines denote the edge of the graphite contact. The Chern numbers of the domains are indicated in (a).